\documentclass[graybox]{svmult}

\usepackage{mathptmx}       
\usepackage{helvet}         
\usepackage{courier}        
\usepackage{type1cm}        
\usepackage{makeidx}         
\usepackage{graphicx}        
\usepackage{multicol}        
\usepackage[bottom]{footmisc}

\usepackage{amsmath}
\usepackage{subfigure}

\usepackage{tabularx}
\usepackage[table]{xcolor}
\definecolor{forestgreen}{HTML}{228B22}
\definecolor{indigo}{HTML}{4B0082}

\makeindex             

\begin{document}

\title*{Evolution of the Global Risk Network Mean-Field Stability Point}
\author{Xiang Niu, Alaa Moussawi, Noemi Derzsy, Xin Lin, Gyorgy Korniss and Boleslaw K. Szymanski}
\institute{Xiang Niu \at Network Science and Technology Center Rensselaer Polytechnic Institute, Troy, NY 12180, USA, \email{nx.niuxiang@gmail.com}
\and Alaa Moussawi \at Network Science and Technology Center Rensselaer Polytechnic Institute, Troy, NY 12180, USA, \email{alaamoussawi@gmail.com}
\and Noemi Derzsy \at Network Science and Technology Center Rensselaer Polytechnic Institute, Troy, NY 12180, USA, \email{noemiderzsy@gmail.com}
\and Xin Lin \at Network Science and Technology Center Rensselaer Polytechnic Institute, Troy, NY 12180, USA, \email{linxin506@gmail.com}
\and Gyorgy Korniss \at Network Science and Technology Center Rensselaer Polytechnic Institute, Troy, NY 12180, USA, \email{korniss@rpi.edu}
\and Boleslaw K. Szymanski \at Network Science and Technology Center Rensselaer Polytechnic Institute, Troy, NY 12180, USA, \email{szymab@rpi.edu}}
%
%
\maketitle
\newcolumntype{C}[1]{>{\centering\arraybackslash} m{#1}}

\abstract*{With a steadily growing human population and rapid advancements in technology, the global human network is increasing in size and connection density. This growth exacerbates networked global threats and can lead to unexpected consequences such as global epidemics mediated by air travel, threats in cyberspace, global governance, etc. A quantitative understanding of the mechanisms guiding this global network is necessary for proper operation and maintenance of the global infrastructure. Each year the World Economic Forum publishes an authoritative report on global risks, and applying this data to a CARP model, we answer critical questions such as how the network evolves over time. In the evolution, we compare not the current states of the global risk network at different time points, but its steady state at those points, which would be reached if the risk were left unabated. Looking at the steady states show more drastically the differences in the challenges to the global economy and stability the world community had faced at each point of the time. Finally, we investigate the influence between risks in the global network, using a method successful in distinguishing between correlation and causation. All results presented in the paper were obtained using detailed mathematical analysis with simulations to support our findings.}

\abstract{With a steadily growing human population and rapid advancements in technology, the global human network is increasing in size and connection density. This growth exacerbates networked global threats and can lead to unexpected consequences such as global epidemics mediated by air travel, threats in cyberspace, global governance, etc. A quantitative understanding of the mechanisms guiding this global network is necessary for proper operation and maintenance of the global infrastructure. Each year the World Economic Forum publishes an authoritative report on global risks, and applying this data to a CARP model, we answer critical questions such as how the network evolves over time. In the evolution, we compare not the current states of the global risk network at different time points, but its steady state at those points, which would be reached if the risk were left unabated. Looking at the steady states show more drastically the differences in the challenges to the global economy and stability the world community had faced at each point of the time. Finally, we investigate the influence between risks in the global network, using a method successful in distinguishing between correlation and causation. All results presented in the paper were obtained using detailed mathematical analysis with simulations to support our findings.}

\section{Introduction}
Recently, cascading failures have been extensively studied~\cite{szymanski2015failure, moussawi2017limits, lin2017limits,  watts2002simple, asztalos2012distributed, roukny2013default}. Most studies focus on financial institutions, internet and infrastructure systems~\cite{dobson2007complex, haldane2011systemic}.
Cascades of global risks can be modeled with Cascades Alternating Renewal Processes (CARP)~\cite{cox1977theory}. In the simplest version of this model, a system alternates between states of being {\it operational} or in {\it failure}, denoted by 0 and 1 respectively. In terms of global risks, an active risk is represented by a failed node (state 1), and a dormant risk is a node in the operational state (state 0). During state transitions, we assume the risks are triggered by non-homogeneous Poisson processes~\cite{szymanski2015failure}. Given the complexity of real-world interactions, global risk probabilities have strong interdependence.

As mentioned above, the CARP models entities oscillating between a set of states. At the minimum, which is the case of global risk network analyzed in this paper, two states are defined as operational and failed. In global risk network, each risk is in either in active or dormant state and at random times undergoes one of the two transitions. The first is risk activation that moves the risk from dormant to active state and the other is recovery transition which reverses the first one. In~\cite{lin2017limits}, we show a model using three states in which recovery is a state not a transition. Here, we further subdivide the activation transition into internal activation and external activation transitions. The three transitions are represented as Poisson processes defined as follows~\cite{szymanski2015failure}.

\begin{itemize}
	\item Internal activation: a dormant risk $i$ is activated internally with intensity $\lambda_i^{int}$. The Poisson probability of transition in time unit is $p_i^{int}=1-e^{-\lambda_i^{int}}$.
	\item External activation: a dormant risk $i$ is activated externally by an active risk $j$ with intensity $\lambda_{i}^{ext}$. The corresponding Poisson probability is $p_{ji}^{ext}=1-e^{-\lambda_{i}^{ext}}$.
	\item Internal recovery: an active risk $i$ is deactivated internally with probability $p_{i}^{rec}$. Conversely, the active risk $i$ continues being active with intensity $\lambda_{i}^{con}$. The corresponding Poisson probability is $p_{i}^{con}=1-e^{-\lambda_{i}^{con}}=1-p_{i}^{rec}$.
\end{itemize}

There are some similarities between the CARP model~\cite{cox1977theory} and epidemic models, such as SIS~\cite{epidemic}, if we consider risks as a population undergoing infection with the activation pathogen. Yet deeper comparison reveals that the CARP model is more complex by including hidden from observations exogenous (infection) and endogenous (falling into activity without or not through contact with active risks) transitions into activity. Thus, the challenges like finding model parameters matching historical data to enable the model to compute probabilities of exogenous and endogenous activations and establish probabilities for steady states are not present in epidemics models. Another significant difference is the transition probabilities evolve as new threats arise, old ones die, and some existing risks increase their probability to activate, while others decrease their probabilities as a result of the increasing resilience mounted by threatened governments, organizations and people. 

Based on the likelihood $l_i$ provided by experts in WEF Global Risk Reports~\cite{WEF2013, WEF2017} for each risk $i$, we obtain a normalized likelihood $L_i$, which indicates how likely a risk $i$ is to be active. Let $\lambda_i^{int}=-\alpha\ln(1-L_i)$, $\lambda_i^{ext}=-\beta\ln(1-L_i)$, $\lambda_i^{con}=-\gamma\ln(1-L_i)$, Szymanski, et al.~\cite{szymanski2015failure} define
\begin{eqnarray}
p_i^{int}&=&1-(1-L_i)^{\alpha} \nonumber\\
p_{ji}^{ext}&=&1-(1-L_i)^{\beta} \nonumber\\
p_i^{con}&=&1-(1-L_i)^{\gamma}  .
\label{eq_poisson}
\end{eqnarray}
The advantage of Eq.~(\ref{eq_poisson}) is that the probabilities of the three Poisson processes are all defined by the normalized likelihood $L_i$ with additional (positive) control parameters $\alpha, \beta, \gamma$. The quality of WEF expert likelihood assessment can be measured by how close $\gamma$ expressed in time units used by expert is to one, since with $\gamma$=1, the probability of an active risk $i$ to continue being active is equal to the normalized assessment $L_i$.

After combining the probabilities of all possible Poisson processes in the risk network, we obtain the state transition probabilities~\cite{szymanski2015failure} as
\begin{eqnarray}
P_i(t)^{0\rightarrow1}&=&1-(1-p_i^{int})(1-p_{ji}^{ext})^{\sum_{j\in N_i} s_j(t)} \nonumber\\
P_i(t)^{1\rightarrow1}&=&p_i^{con} \nonumber\\
P_i(t)^{0\rightarrow0}&=&1-P_i(t)^{0\rightarrow1} \nonumber\\
P_i(t)^{1\rightarrow0}&=&1-P_i(t)^{1\rightarrow1}   , 
\label{eq_transition}
\end{eqnarray}
where $P_i(t)^{0\rightarrow1}$ represents the probability of the state of risk $i$ being changed from dormant to active at time $t$, $N_i$ is the set of nodes connected to risk $i$, and $s_j(t)$ represents the state of risk $j$ at time $t$ (0 is dormant, 1 is active).

The transition probability of risk $i$ at time $t$ is given as $P_i(t)^{s_i(t)\rightarrow s_i(t+1)}$, the vector of transition probabilities of all risks is $\overrightarrow{S(t)}=\prod_{i=1}^{R} P_i(t)^{s_i(t)\rightarrow s_i(t+1)}$, where $R$ denotes the number of risks. The likelihood of the sequence of state transition of all risks is $L(\overrightarrow{S(1)},\overrightarrow{S(2)},...,\overrightarrow{S(T)})=\prod_{t=1}^{T-1} \prod_{i=1}^{R} P_i(t)^{s_i(t)\rightarrow s_i(t+1)}$, $T$ representing the number of time steps during the entire model simulation~\cite{szymanski2015failure}. The log-likelihood is 
\begin{eqnarray}
\ln L(\overrightarrow{S(1)},\overrightarrow{S(2)},...,\overrightarrow{S(T)})=\sum_{t=1}^{T-1} \sum_{i=1}^{R} \ln P_i(t)^{s_i(t)\rightarrow s_i(t+1)}   .
\label{eq_mle}
\end{eqnarray}

We compute the optimal values of model parameters $\alpha, \beta$, and $\gamma$ by maximizing the loglikelihood in
Eq.~(\ref{eq_mle}) of the observed state transition processes~\cite{dempster1977maximum, pawitan2001all}. Then, with these optimal values, we can simulate CARP model to collect the changes of the stable states of the global risk network at any point of time for which we have historical data.

From the observations of both historical events and WEF Global Risk Reports~\cite{WEF2013, WEF2017}, we find that the global risk network constantly evolves as new threats arise, and some existing risk increase their probability to activate, while others decline their probabilities as a result of the increasing resilience mounted by threatened governments, organizations and people. This evolution causes continuous changes in the current state of the global risks and their probabilities. However, if left unabated, the global risk network would move from the current state to the steady state, in which some risk activities will be much more frequent and their threats much more pronounced. This is why, in this paper, we do not compare the states of the global risk network at some points of time (in case of this paper, January 2013 and August 2017). Instead, we compare the steady states to which these initial states would move if no changes to the system had been introduced. Looking at the steady states show more drastically the differences in the challenges to the global economy and stability the world community had faced at each point of the time.

\begin{figure}
	\centering
	\includegraphics[width=1.5\textwidth, angle =-90]{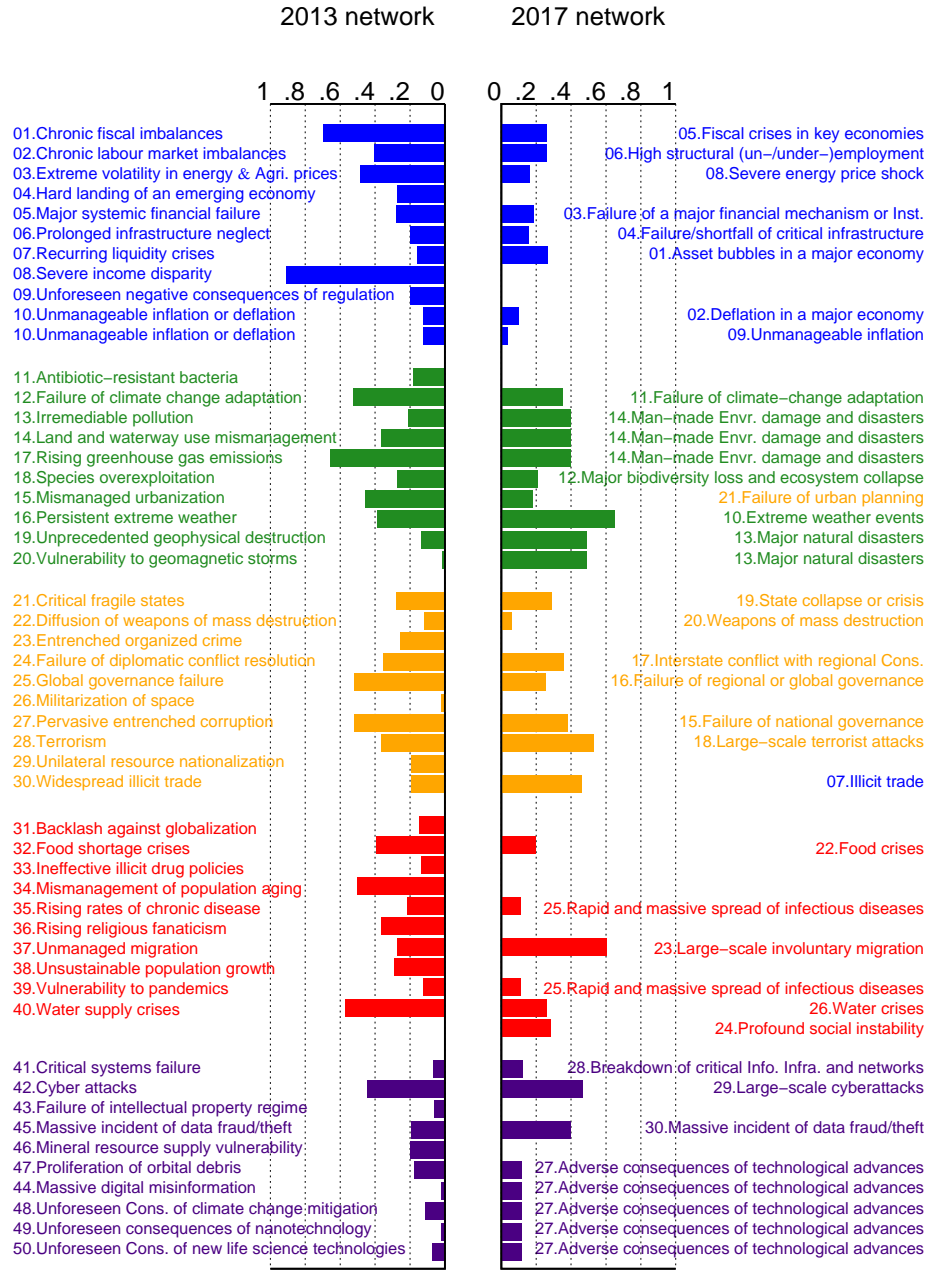}
	\caption{The mean-field stable probabilities of risks being active in 2013 and 2017 networks.}
	\label{fig_steady_state}
\end{figure}

\section{Risk Networks}
In the World Economic Forum (WEF) Global Risk Reports~\cite{WEF2013, WEF2017}, experts define risks grouped into 5 categories. The list of risks is shown in Fig.~\ref{fig_steady_state}. In the 2013 risk network, 1-10 are economic risks, 11-20 are environmental risks, 21-30 are geopolitical risks, 31-40 are societal risks, and 41-50 are technological risks. In the 2017 risk network, 1-9 are economic risks, 10-14 are environmental risks, 15-20 are geopolitical risks, 21-26 are societal risks, and 27-30 are technological risks.

We utilize the event dataset created for~\cite{szymanski2015failure}, which includes news, academic articles, Wikipedia entries, etc. from Jan. 2000 to Dec. 2012, and based on this we generate 13x12x50=7800 data points for the 2013 risk network. For the 2017 risk network, we relabel prior events and collect new events from Jan. 2013 to Aug. 2017. Thus the total number of data points is (17x12+8)x30=6360. Each data point indicates if a risk is active or dormant in a certain month.

We obtain the optimal values of model parameters $\alpha, \beta, \gamma$ in Table~\ref{table_network_property} by maximizing the log-likelihood from Eq.~(\ref{eq_mle}) of the observed state transition processes~\cite{dempster1977maximum, pawitan2001all}. The state transition processes are expressed in time measured in monthly units. In the 2013 network, most of the event-triggered risks are continually active or dormant. In the 2017 network many risks are intermittently active, thus both risk activation and recovery probabilities increase (in terms of parameters $\alpha, \beta$ increase, $\gamma$ decreases). Although the average degree of the 2013 risk network is larger than that of the 2017 risk network, since the 2017 network is smaller, its risks have a higher probability of being connected, and have larger mean clustering coefficient and smaller diameter.

\begin{table}
\caption{Network properties of 2013 and 2017 risk networks.}
\centering
\begin{tabular}{|c|c|c|c|c|c|c|c|c|c|} \hline
Network & Nodes & Edges & Avg. degree & Edge Prob. & Avg. CC. & Diameter & $\alpha$ & $\beta$ & $\gamma$ \\ \hline
2013 & 50 & 515 & 20.60 & 0.42 & 0.61 & 3 & $3.04e^{-3}$ & $1.17e^{-3}$ & 3.56 \\ \hline
2017 & 30 & 275 & 18.33 & 0.63 & 0.74 & 2 & $5.28e^{-3}$ &  $3.03e^{-3}$ & 2.50 \\ \hline
\end{tabular}
\label{table_network_property}
\end{table}

\subsection{Stability Points}

\begin{figure*}
	\centering
	\includegraphics[width=\textwidth]{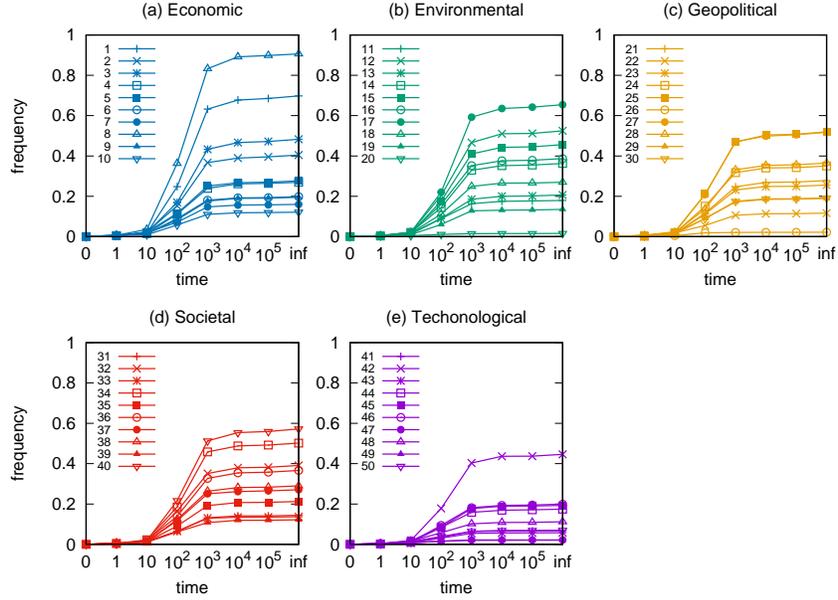}
	\caption{Asymptotic mean-field stable frequencies of selected risks being active are shown for 2013 risk network. Each frequency value is averaged over 1000 runs. For time denoted on x axis as Inf, the frequencies are calculated by Eq.~(\ref{eq_risk_probability}).}
	\label{fig_cascade}
\end{figure*}

With the learned parameters $\alpha, \beta, \gamma$, and the activation and recovery probabilities, we perform Monte Carlo simulations of the cascades of global risks. Fig.~\ref{fig_cascade} shows the frequency of a risk being active at each time step $t$ during the simulation of 2013 risk network. The frequency of risk $i$ being active at time $t$ is the number of simulations in which risk $i$ is active as a fraction of all simulations.
We find that the frequency distributions of risks being active eventually saturates. By denoting the probability of risk $i$ being active at time $t$ as $p_i(t)$, we define such frequency to be stable when $p_i(t) \approx p_i(t+1)$. By plugging in the state transition probabilities in Eq.~(\ref{eq_transition}), we have
\begin{equation}
(1-p_i(t))P_i(t)^{0\rightarrow1}+p_i(t)P_i(t)^{1\rightarrow1}=p_i(t+1)=p_i(t)  .\\
\label{eq_probability_transition}
\end{equation}
Thus, 
\begin{eqnarray}
\hat{p_i}&=&\frac{P_i^{0\rightarrow1}}{P_i^{0\rightarrow1}+1-P_i^{1\rightarrow1}}=\frac{1-(1-L_i)^{\alpha+\beta\sum_{j\in N_i}\hat{p_j}}}{1-(1-L_i)^{\alpha+\beta\sum_{j\in N_i}\hat{p_j}}+(1-L_i)^{\gamma}}   ,
\label{eq_risk_probability}
\end{eqnarray}
where $\hat{p_i}$ is the steady state  probability of risk $i$ being active, computed with a successive approximation method (Fig.~\ref{fig_steady_state}). 



\subsection{Risk Evolution}

Fig.~\ref{fig_steady_state} shows the evolution of the global risk network and its mean-field stability point. To clearly see the changes in risks, we display related risks and their indices in the two networks side by side. From 2013 to 2017 five risk categories remain the same, while around 20 risks vanish or merge into other risks. Risk 10, ``Unmanageable inflation or deflation" splits into risk 02 ``Deflation in a major economy" and 09 ``Unmanageable inflation". Risk 15 ``Mismanaged urbanization" changes from Environmental to Geopolitical and is renamed as risk 21 ``Failure of urban planning". Risk 30 ``Widespread illicit trade" is changed from Geopolitical to Economic and renamed as risk 07 ``Illicit trade", whereas Risk 24 ``Profound social instability" is newly proposed.

Comparing the mean-field stable probabilities of risks from 2013 and 2017, we find that the probabilities of economic risks generally and significantly decrease, reflecting a gradual global recovery from the 2008 economic crisis. Only 07 ``Illicit trade", 18 ``Large scale terrorist attacks" and 23 ``Large scale involuntary migration" significantly increase in the category of economic, geopolitical, and societal risks, respectively. This reveals the negative effects of globalization. Considering environmental risks, we find that man made risks decrease, while only nature risks 10 ``Extreme weather events" and 13 ``Major natural disasters" increase. The abilities of the public to prevent the environmental degradation are improved. In technological risks category, risk 30 ``Massive incident of data fraud/theft" increases due to the boom of private data in the internet era.

\begin{figure}
	\centering
	\includegraphics[width=\textwidth]{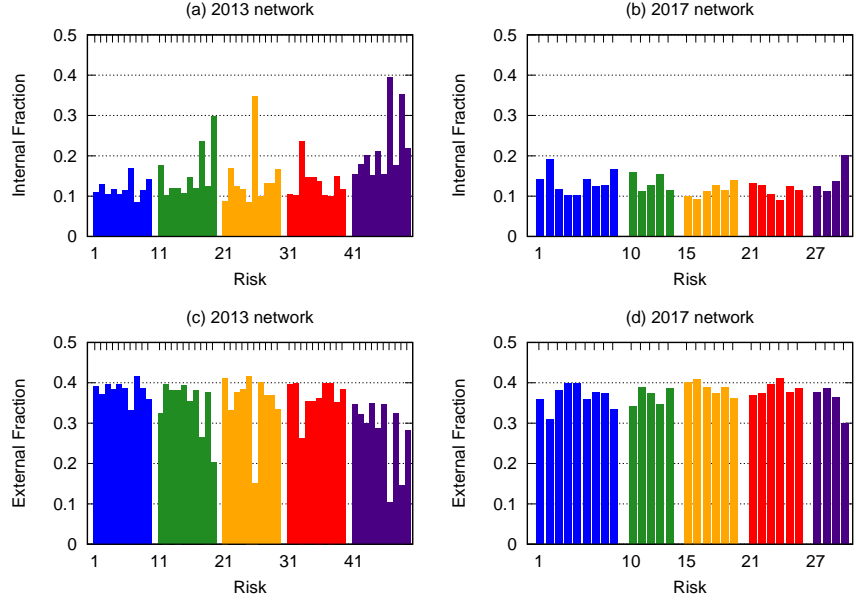}
	\caption{The fraction of internal and external activations for each risk.}
	\label{fig_transition}
\end{figure}

\section{Transition Fractions}
With the steady state probability of a risk being active, we can compute the probability of three different transitions:
\begin{itemize}
	\item internal activation: $A_i^{int}=(1-\hat{p_i})p_i^{int}$ probability of dormant risk $i$ $(1-\hat{p_i})$ being triggered internally $p_i^{int}$.
	\item external activation: $A_i^{ext}=(1-\hat{p_i})[1-(1-p_{ji}^{ext})^{\sum_{j\in N_i}\hat{p_j}}]$ probability of dormant risk $i$ $(1-\hat{p_i})$ being triggered externally $1-(1-p_{ji}^{ext})^{\sum_{j\in N_i}\hat{p_j}}$.
	\item internal recovery: $A_i^{rec}=\hat{p_i}p_i^{rec}$ probability of active risk $i$ $(\hat{p_i})$ recovering $p_i^{rec}$.
\end{itemize}
For simplicity, we ignore the probability of a risk being activated both internally and externally with probability $(1-\hat{p_i})p_i^{int}[1-(1-p_{ji}^{ext})^{\sum_{j\in N_i}\hat{p_j}}]$ (the value is negligible). Thus, the three transitions can be treated as independent variables. With the probabilities of transitions, we can get the fraction of one transition to all possible transitions for each risk by setting $a_i^{int}=\frac{A_i^{int}}{A_i^{int}+A_i^{ext}+A_i^{rec}}$, $a_i^{ext}=\frac{A_i^{ext}}{A_i^{int}+A_i^{ext}+A_i^{rec}}$, $a_i^{rec}=\frac{A_i^{rec}}{A_i^{int}+A_i^{ext}+A_i^{rec}}$. The values are plotted in Fig.~\ref{fig_transition}. By nature of activation and recovery, for each risk, the recovery accounts for half of the transitions. For most of the risks, external activation is more frequent than internal activation. Around one fourth of the activations are triggered internally, while the other three fourths are triggered externally. According to Eq.~(\ref{eq_poisson})

\begin{eqnarray}
\frac{A_i^{ext}}{A_i^{int}}=\frac{[1-(1-p_{ji}^{ext})^{\sum_{j\in N_i}\hat{p_j}}]}{p_i^{int}}=\frac{1-(1-L_i)^{\beta\sum_{j\in N_i}\hat{p_j}}}{1-(1-L_i)^{\alpha}}.
\end{eqnarray}
Applying Taylor's Approximation, $(1-x)^m\approx 1-mx$, when $mx\to0$, we obtain $\frac{1-(1-L_i)^{\beta\sum_{j\in N_i}\hat{p_j}}}{1-(1-L_i)^{\alpha}}\approx\frac{L_i\beta\sum_{j\in N_i}\hat{p_j}}{L_i\alpha}=\frac{\beta\sum_{j\in N_i}\hat{p_j}}{\alpha}$. The average of $\frac{\beta\sum_{j\in N_i}\hat{p_j}}{\alpha}$ is around 2.69 for the 2013 network and 3.07 for the 2017 network.
Despite $\beta$ being smaller than $\alpha$, the network structure amplifies the effects of external activations.

The risks in the 2017 network tend to cluster together and are more likely to be triggered externally in general. A few risks in the 2013 network are almost isolated and are always triggered internally, including 20 ``Vulnerability to geomagnetic storms" (6 neighbors), 26 ``Militarization of space" (5 neighbors), 47 ``Proliferation of orbital debris" (3 neighbors) and 49 ``Unforeseen consequences of nanotechnology" (2 neighbors).



\section{Risk Influence}
In this section, we calculate the influence of risks on each other. In the experiments, we first disable a risk $i$ by setting its normalized likelihood $L_i=0$, and then calculate the new external activation frequency of risk $j$ as $a_{j-i}^{ext}$ ($j\neq i$). We obtain
\begin{equation}
I_{i\rightarrow j}=a_j^{ext}-a_{j-i}^{ext},
\end{equation}
where $I_{i\rightarrow j}$ is an indicator of the influence risk $i$ has on risk $j$, quantifying the external activation effects of risk $i$ onto risk $j$.

\begin{figure}
	\centering
	\sidecaption
	\includegraphics[scale=0.5]{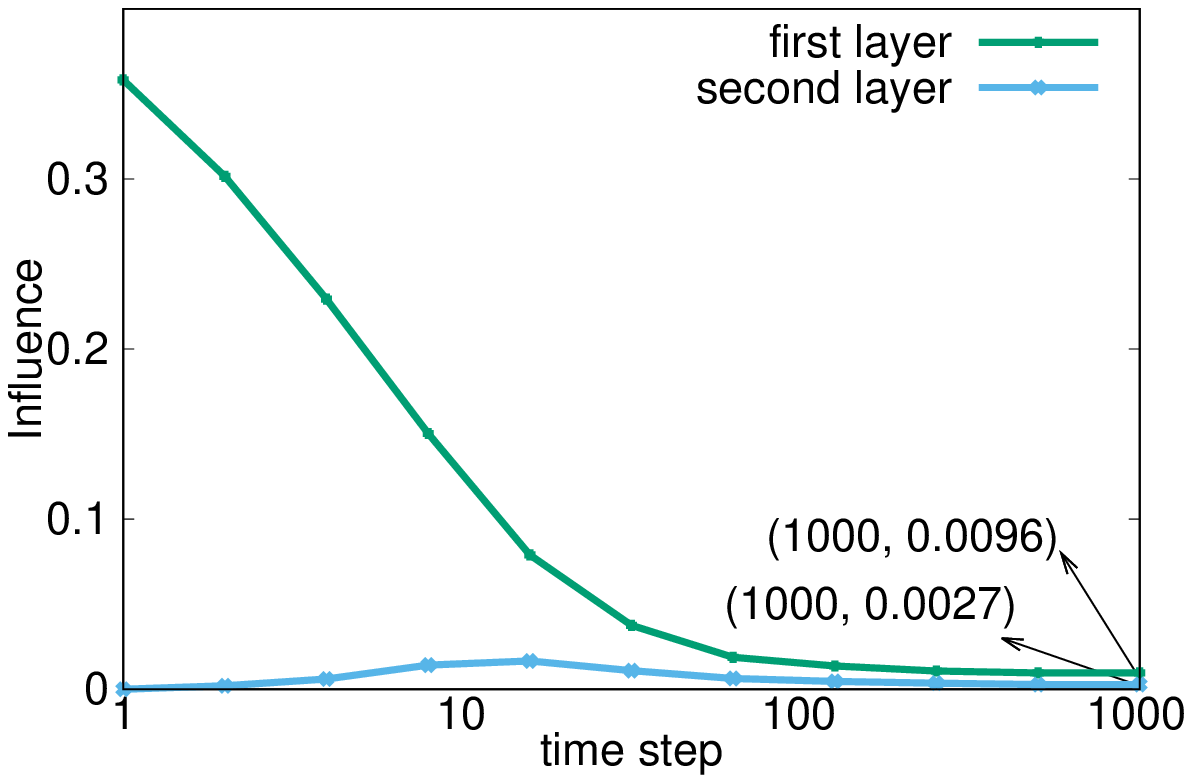}
	\caption{Monte Carlo simulation of risk influences. For each risk $i$, we calculate separately the average influence on the risks in its first and second neighborhood layers. We activate the risk $i$ at step 0 and record its influences at every time step. Each influence value in the plot is averaged over 1000 runs and 30 risks.}
	\label{fig_influence_difference_direct_all}
\end{figure}

The influence of risk 01 ``Asset bubbles in a major economy" in the 2017 network is most pronounced of economic risks, with the top three influenced risks being economic. The top sixteen influenced risks are simply the nearest neighbors of the source risk. When testing the influence of other risks in both the 2013 and the 2017 networks, we found that for any risk $i$ with $n$ nearest neighbors, its top $n$ influenced risks are always those neighbors. The topology of the network has a prominent effect on the influence.
Fig.~\ref{fig_influence_difference_direct_all} shows an example of the difference of influences of risks on their one-hop (nearest) and two-hops neighbors in 2017 network. One-hop neighbors are immediately affected, but the impact decreases with time; while two-hop neighbors need more time to react, with a peak at approximately 20 time steps.

\begin{figure}
	\centering
	\includegraphics[width=\textwidth]{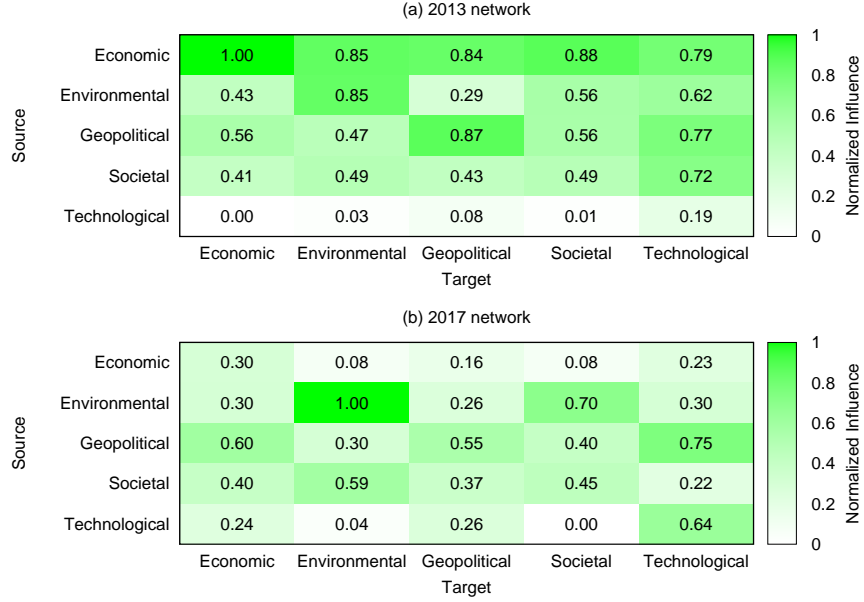}
	\caption{Normalized influences among risk categories.}
	\label{fig_category_influence}
\end{figure}

Fig.~\ref{fig_category_influence} shows the influence of a category of risks on other categories, which discerns between cause and correlation of risks. The influence of category $c_i$ on $c_j$ is calculated as 
\begin{eqnarray}
I_{c_i\rightarrow c_j}=\left\{
			\begin{array}{ll}
			\frac{\sum_{i\in c_i, j\in c_j}I_{i\rightarrow j}}{|c_i||c_j|} (c_i\neq c_j)\\
			\frac{\sum_{i\in c_i, j\in c_j, j\neq i}I_{i\rightarrow j}}{|c_i|(|c_j|-1)} (c_i= c_j)
			\end{array}
		\right. ,
\end{eqnarray}
where $I_{i\rightarrow j}$ is the influence of risk $i$ on $j$, $c_i, c_j$ represent one of the 5 risk categories: economic, environmental, geopolitical, societal, and technological. With the unity-based normalization of the influences, we find that most categories have large self-influence (diagonal elements). From 2013 to 2017, the most significant changes of risk influences are economic and technological ones. The economic risks used to be the most influential risks and had the highest impact on other risks. However, as of 2017, their influences have decreased. In 2013, technological risks were the most vulnerable risks and had very limited influences on others. Although they are still the least influential risks in 2017, we can see an increasing trend in their influence.


%

\section{Conclusion}
In this paper, we use the CARP model to simulate cascades in global risk networks. With parameters learned by MLE methods from a real event dataset, we compute the mean-field steady state probabilities of risks being active. By computing the difference of external activation frequencies of risk $j$ with enabling and disabling risk $i$, we define an influence $I_{i\rightarrow j}$ of risk $i$ on $j$. The results of the 2013 and 2017 risk networks show significant changes in asymptotic mean-field probabilities of risk activations. The activation probabilities and influences of economic risks are dramatically reduced as a signal of economic recovery since 2013. The increase in activation probability of illicit trade and migration show the negative effects of globalization. Technological risks are becoming influential due to the increase of private data leaks.

\begin{acknowledgement}
This work was supported in part by the Army Research Laboratory under Cooperative Agreement Number W911NF-09-2-0053 (the Network Science CTA), by the Army Research Office grant no. W911NF-16-1-0524, and by DTRA Award No. HDTRA1-09-1-0049. The views and conclusions contained in this document are those of the authors.
\end{acknowledgement}
\bibliographystyle{spmpsci}
\bibliography{risk_cascade}

\end{document}